 \documentclass[showpacs,twocolumn,preprintnumbers,amsmath,amssymb]{revtex4}
 \usepackage{dcolumn}
 \usepackage{bm}

\begin{document}

\preprint{}

\title{Influence of Generalized and Extended Uncertainty Principle on Thermodynamics of
Friedmann-Robertson-Walker universe}

\author {Tao Zhu }\thanks{Email: zhut05@lzu.cn}
\author{Ji-Rong Ren }\thanks{Email: renjr@lzu.edu.cn}
\author{Ming-Fan Li}\thanks{Email: limf07@lzu.cn}

\affiliation{Institute of Theoretical Physics, Lanzhou University,
Lanzhou 730000, P. R. China}

\date{\today}

\begin{abstract}
The influence of the generalized uncertainty principle (GUP) and
extended uncertainty principle (EUP) on the thermodynamics of the
Friedmann-Robertson-Walker (FRW) universe has been investigated. It
is shown that the entropy of the apparent horizon of the FRW
universe gets a correction if one considers the effect of the GUP or
EUP. Moreover, starting with the modified entropy and applying the
first law of thermodynamics, $dE=TdS$, to the apparent horizon of
the FRW universe, we obtain the modified Friedmann equations. The
influence of the GUP or EUP on the thermodynamics of the FRW
universe provides a deep insight into the understanding of the
quantum gravity or large length scale corrections to the dynamics of
the FRW universe.
\end{abstract}



\maketitle

\section{Introduction}
The four thermodynamics laws of black hole, which were originally
derived from the classical Einstein Equation, provide deep insight
into the connection between thermodynamics and Einstein
Equation\cite{Black Hole,ql1}. Recently, this connection has been
investigated extensively in the literatures for Rindler spacetime
and Friedmann-Robertson-Walker (FRW) universe. For Rindler
spacetime\cite{Jacobson}, the Einstein equation can be derived from
the proportionality of entropy to the horizon area, together with
the Clausius relation $\delta Q=TdS$. Here $\delta Q$ and $T$ are
the energy flux and Unruh temperature detected by an accelerated
observer just inside the local Rindler causal horizons through
spacetime point. In FRW universe\cite{gauss-love1}, after replacing
the event horizon of black hole by the apparent horizon of FRW
space-time and assuming that the apparent horizon has an associated
entropy $S$ and temperature $T$
\begin{eqnarray}
S=\frac{A}{4G},~~~~~T=\frac{1}{2\pi \tilde{r}_A},
\end{eqnarray}
one can cast the first law of thermodynamics, $dE=TdS$, to the
Friedmann equations. Here $G$, $A$, and $\tilde{r}_A$ are the
gravitational constant, the area of the apparent horizon, and the
radius of the apparent horizon, respectively. The first law of
thermodynamics not only holds in Einstein gravity, but also in
Guass-Bonnet gravity, Lovelock gravity, and various braneworld
scenarios\cite{fr,gb,brane}. The fact that the first law of
thermodynamics holds extensively in various spacetime and gravity
theories suggests a deep connection between gravity and
thermodynamics. (Some other viewpoints and further developments in
this direction see \cite{padm,mass,cft/frw,ql3,brick,cao2} and
references therein.)

The thermodynamics behavour of spacetime is only one of the features
of Einstein gravity. Another feature is the Hawking radiations at
the event horizon of black holes\cite{hawking} or the apparent
horizon of the FRW spacetime\cite{cao1}. The Hawking radiation is a
quantum mechanics effect in the classical background black hole or
FRW spacetime. Therefore quantum theory, gravitational theory and
thermodynamics meet together at black holes and FRW spacetime. For a
black hole, it radiates and becomes smaller and hotter, finally
disappears when the Hawking radiation ends, leaving behind thermal
radiation described by quantum mechanical mixed states.

However, the analysis of Hawking radiation in the literatures
usually make use of the semi-classical approaches, assuming a
classical background metric and considering a quantum radiation
process. When it comes into the high energy regime, for example a
small black hole whose size can compare with Planck scale or a FRW
universe in the era of Planck time, the effect of quantum gravity
should not be forgotten. In these cases, the conventional
semi-classical approaches are not proper and the complete quantum
theory of gravity is required.

Recently, a growing interest has been focused on the proposal that
the quantum gravity effect might need us turn from the usual
commutation relations of the Heisenberg's uncertainty principle
(HUP) to the generalized uncertainty principle (GUP)\cite{gup1}. The
GUP is a model independent aspect of quantum gravity and can be
derived from different approaches to quantum gravity, such as string
theory\cite{gup3-string}, loop quantum gravity and noncommutative
quantum mechanics\cite{gup2}.

Naturally, one may think that the GUP should influence the
thermodynamics of black holes and FRW universe in the small scale or
in the high energy regime. Indeed, this issue has been investigated
in contexts of black hole physics. As we have known, the GUP affects
the thermodynamics of black holes in two aspects. First, the GUP
might modify the Hawking temperature on the event horizon and may
prevent the total evaporation of a black
hole\cite{gup-hawking,gup-hawking1}. Second, after considering the
GUP, one will get a correction to the Bekenstein-Hawking entropy of
a black hole\cite{gup-entropy1,gup-entropy2}. This correction
modifies the famous entropy-area relation that the entropy of a
black hole is proportional to its area of the event horizon. The
impact of the GUP on other physics systems has also been
investigated extensively, see\cite{other} and references therein.

However, as far as we know, whether the GUP can influence the
thermodynamics of FRW universe is still unknown. Is there indeed a
correction to the entropy on the apparent horizon of the FRW
universe when we consider the effect of the GUP?  If the GUP is
considered,can we still get the Friedmann equations when we apply
the first law of thermodynamics to the apparent horizon? These
problems need to be solved. In this paper, we are going to
investigate these problems. We find that by utilizing the GUP, the
entropy of the apparent horizon of the FRW universe should get a
correction. Moreover, starting with the modified entropy on the
apparent horizon, we will show that the first law of thermodynamics
on the apparent horizon can produce the corresponding modified
Friedmann equations.

However, as a high energy correction to HUP, the GUP should not be
important for the late time FRW universe. In this case, one might
consider the effect of the large length scale modification. (For
example, in Dvali-Gabadadze-Porrati braneworld model, the large
length scale modification to the Einstein gravity on the brane might
lead to the late time acceleration of our universe.) Recently, an
extended uncertainty principle (EUP) has been introduced to
incorporate the effect of the large length
scale\cite{gup-hawking1,eup}. We note here that in this paper, we
adopt the terminology, the extended uncertainty principle (EUP) and
the generalized EUP (GEUP), which were first used in
\cite{gup-hawking1}. Contrary to the GUP, the EUP is the large
length scale correction to the Heisenberg's uncertainty principle.
In black hole physics, the uncertainty principle can be used to
derive the Hawking temperature of Schwarzschild-(anti)de Sitter
black hole\cite{gup-hawking1,eup1}. Like in the case of the GUP, it
is also of interest to investigate the impact of the EUP on the
thermodynamics of the FRW universe. In this paper, we will also
consider this issue.

Therefore, the organization of this paper is as follows. In Section
II, we investigate the influence of the GUP on the thermodynamics of
the FRW universe, and in Section III, we generalize the discussions
of the GUP to the EUP case. The Section IV are our summary and
discussions. Throughout the paper, the units $c\equiv\hbar\equiv
k_B\equiv1$ are used.

\section{The GUP case}
Let us begin with the GUP, which is usually given by\cite{gup1}
\begin{eqnarray}
\delta x\delta p\geq 1+\alpha^2l_p^2 \delta p^2,\label{GUP}
\end{eqnarray}
where $l_p$ is the Planck length, and $\alpha$ is a dimensionless
real constant. The GUP has an immediate consequence that there is a
minimal length with the Planck scale,
\begin{eqnarray}
\delta x\geq \frac{1}{\delta p}+\alpha^2l_p^2 \delta p \geq 2|\alpha
l_p|.
\end{eqnarray}
This minimal length characterizes the absolute minimum in the
position uncertainty.

Another consequence of GUP is the modified momentum uncertainty.
After some simple manipulations, the momentum uncertainty can be
written as
\begin{eqnarray}
\delta p\geq\frac{1}{\delta x}[\frac{\delta x^2}{2\alpha^2
l_p^2}-\frac{ \delta x^2}{2\alpha^2 l_p^2}
\sqrt{1-\frac{4\alpha^2l_p^2}{(\delta x)^2}}]=\frac{1}{\delta
x}f_G(\delta x^2),\label{gup3}
\end{eqnarray}
where
\begin{eqnarray}
f_G(\delta x^2)=\frac{\delta x^2}{2\alpha^2 l_p^2}-\frac{ \delta
x^2}{2\alpha^2 l_p^2} \sqrt{1-\frac{4\alpha^2l_p^2}{(\delta
x)^2}}\end{eqnarray} characterizes the departure of the GUP from the
Heisenberg uncertainty principle $\delta p\geq 1/\delta x$.

We consider a ($n+1$)-dimensional FRW universe, whose linear element
is given by
\begin{eqnarray}
ds^2=-dt^2+a^2(\frac{dr^2}{1-kr^2}+r^2d\Omega_{n-1}^2),
\end{eqnarray}
where $d\Omega_{n-1}^2$ denotes the line element of an
($n-1$)-dimensional unit sphere, $a$ is the scale factor of our
universe and $k$ is the spatial curvature constant. In FRW
spacetime, there is a dynamical apparent horizon, which is a
marginally trapped surface with vanishing expansion. Using the
notion $\tilde{r}=ar$, the radius of the apparent horizon can be
written as
\begin{eqnarray}
\tilde{r}_A=\frac{1}{\sqrt{H^2+k/a^2}},
\end{eqnarray}
where $H$ is the Hubble parameter, $H\equiv \dot{a}/a$ (the dot
represents derivative with respect to the cosmic time $t$). On the
apparent horizon, if we suppose that the apparent horizon has an
associated entropy $S$ and temperature $T$
\begin{eqnarray}
S=\frac{A}{4G},~~~~~T=\frac{1}{2\pi \tilde{r}_A},
\end{eqnarray}
(where $A$ is the apparent horizon area
$A=n\Omega_n\tilde{r}_A^{n-1}$ with
$\Omega_n=\pi^{n/2}/\Gamma(n/2+1)$ being the volume of an
$n$-dimensional unit sphere.) it has been confirmed
\cite{gauss-love1} that the first law of thermodynamics,
\begin{eqnarray}
dE=TdS,
\end{eqnarray}
can reproduce the Friedmann equations
\begin{eqnarray}
\dot{H}-\frac{k}{a^2}=-\frac{8\pi G}{n-1}(\rho+p),\\
H^2+\frac{k}{a^2}=\frac{16\pi G}{n(n-1)}\rho.\label{fr}
\end{eqnarray}
Here $\rho$ is the energy density of cosmic fluid and $dE=d(\rho V)$
is the energy flow pass through the apparent horizon. Note that in
order to get Eq.(\ref{fr}), one should use
\begin{eqnarray}
\dot{\rho}+n H(\rho+p)=0,
\end{eqnarray}
which is the continuity (conservation) equation of the perfect
fluid.

Now we consider the impact of the GUP on thermodynamics of FRW
universe. We consider the case that the apparent horizon having
absorbed or radiated a particle with energy $dE$. As point out
in\cite{gup-hawking1}, one can identify the energy of the absorbed
or radiated particle as the uncertainty of momentum,
\begin{eqnarray}
dE\simeq\delta p.
\end{eqnarray}
By considering the quantum effect of the absorbed or radiated
particle, which implies the Heisenberg uncertainty principle $\delta
p\geq \hbar/\delta x$, the increase or decrease in the area of the
apparent horizon can be expressed as
\begin{eqnarray}
dA=\frac{4G}{T}dE\simeq \frac{4G}{T}\frac{1}{\delta x}.\label{area1}
\end{eqnarray}
In the above we didn't consider the impact of the GUP. When the
effect of the GUP (\ref{gup3}) is considered, the change of the
apparent horizon area can be modified as
\begin{eqnarray}
dA_G=\frac{4G}{T}dE\simeq\frac{4G}{T}\frac{1}{\delta x}f_G(\delta
x^2).\label{area2}
\end{eqnarray}
Using Eq.(\ref{area1}), we have
\begin{eqnarray}
dA_G=f_G(\delta x^2)dA.\label{area3}
\end{eqnarray}
Take into account that the position uncertainty $\delta x$ of the
absorbed or radiated particle can be chosen as its Compton length,
which has the order of the inverse of the Hawking temperature, one
can take\cite{gup-entropy1}
\begin{eqnarray}
\delta x\simeq 2\tilde{r}_A=2(\frac{A}{n\Omega_n})^{\frac{1}{n-1}}.
\end{eqnarray}
Thus, the departure function $f_G(\delta x^2)$ can be re-expressed
in terms of $A$,
\begin{eqnarray}
f_G(A)=\frac{2}{\alpha^2l_p^2}(\frac{A}{n\Omega_n})^{\frac{2}{n-1}}(1-\sqrt{1-\alpha^2l_p^2(\frac{n\Omega_n}{A})^{\frac{2}{n-1}}}).
\end{eqnarray}
Here and hereafter we use $f_G(A)$ represent the departure function
$f_G(\delta x^2)$. At $\alpha=0$, we express $f_G(A)$ by Taylor
series
\begin{eqnarray}
f_G(A)&=&1+\frac{\alpha^2l_p^2}{4}(\frac{n\Omega_n}{A})^{\frac{2}{n-1}}+
\frac{(\alpha^2l_p^2)^2}{8}(\frac{n\Omega_n}{A})^{\frac{4}{n-1}}\nonumber\\
&+&\sum_{d=3}c_d(\alpha
l_p)^{2d}(\frac{n\Omega_n}{A})^{\frac{2d}{n-1}},\label{f}
\end{eqnarray}
where $c_d$ is a constant.

If we substitute (\ref{f}) into Eq.(\ref{area3}) and integrating, we
can get the modified area $A_G$ from the GUP. Then we can also get
the correction to the entropy area relation by using $S_G=A_G/4G$.
But integrating Eq.(\ref{area3}) might be complicated and
dimensional dependent. Therefore, we should divide our discussions
into three cases: (1)$n=3$; (2)$n>3$ and $n$ is a even number;
(3)$n>3$ and $n$ is a odd number.

\subsection{The $n=3$ case}
When $n$=3, we have
\begin{eqnarray}
f_G(A)&=&1+\pi\alpha^2l_p^2\frac{1}{A}+
2(\pi \alpha^2l_p^2)^2\frac{1}{A^2}\nonumber\\
&+&\sum_{d=3}c_d(4\pi\alpha^2 l_p^2)^{2d}\frac{1}{A^d}.\label{f1}
\end{eqnarray}
Substituting (\ref{f1}) into (\ref{area3}) and integrating, we
obtain
\begin{eqnarray}
A_G&=&A+\pi\alpha^2l_p^2lnA-2(\pi\alpha^2l_p^2)^2\frac{1}{A}\nonumber\\
&-&\sum_{d=3}\frac{c_d(4\pi\alpha^2
l_p^2)^{2d}}{d-1}\frac{1}{A^{d-1}}+c
\end{eqnarray}
where $c$ is the integral constant. By making use of
Bekenstein-Hawking area law, $S=A/4G$, we can obtain the expression
of the entropy of the apparent horizon including the effect of the
GUP. That is, the modified entropy is given by
\begin{eqnarray}
S_G&=&\frac{A}{4G}+\frac{\pi\alpha^2l_p^2}{4G}ln\frac{A}{4G}-2(\frac{\pi\alpha^2l_p^2}{4G})^2(\frac{A}{4G})^{-1}\nonumber\\
&-&\sum_{d=3}\frac{c_d(\frac{16\pi^2\alpha^4l_p^4
}{4G})^{d}}{d-1}(\frac{A}{4G})^{1-d}+const.\label{entropy1}
\end{eqnarray}
This relation has the standard form of the entropy-area relation as
given by other approaches in black holes\cite{log1,log2,other1}. The
point which should be stressed here is that the coefficient of the
logarithmic correction term is positive. This is different with the
results in Refs.\cite{log1,log2}.  As pointed out in some
literatures\cite{log2}, the coefficient of the logarithmic
correction term is controversial. Our result shows that the
correction to the entropy from the GUP gives an opposite
contribution to the area entropy.

Recently, starting with a modified entropy-area relation, Cai, Cao
and Hu \cite{cao2} have shown that the first law of thermodynamics
on the apparent horizon can produce a modified Friedmann equation.
Now we give the main results of Cai, Cao and Hu's approach and apply
their approach to the case of the modified entropy-area relation
(\ref{entropy1}). Suppose the apparent horizon has an entropy
$S_G(A)$. Applying the first law of thermodynamics to the apparent
horizon of FRW universe, we can obtain the corresponding Friedmann
equations
\begin{eqnarray}
(\dot{H}-\frac{k}{a})S'_G(A)=-\pi(\rho+p),\label{fridemann1}\\
\frac{8\pi G}{3}\rho=-\frac{\pi}{G}\int
S'_G(A)(\frac{4G}{A})^2dA,\label{fridemann2}
\end{eqnarray}
where a prime stands for the derivative with respect to $A$.
Eq.(\ref{fridemann1}) and (\ref{fridemann2}) are nothing but the
modified first and second Friedmann equation corresponding to the
modified apparent horizon entropy $S_G(A)$.

Noticing that $S_G=A_G/4G$ and considering Eq.(\ref{area3}), we can
obtain
\begin{eqnarray}
S'_G(A)=\frac{f_G(A)}{4G}.\label{entropy x}
\end{eqnarray}
Substituting (\ref{f1}) and (\ref{entropy x}) into the modified
Friedmann equation (\ref{fridemann1}) and (\ref{fridemann2}), we can
obtain the modified Friedmann equations after considering the GUP,
that is
\begin{eqnarray}
(\dot{H}-\frac{k}{a})[1+\pi\alpha^2l_p^2\frac{1}{A}+
2(\pi \alpha^2l_p^2)^2\frac{1}{A^2}\nonumber\\
+\sum_{d=3}c_d(4\pi\alpha^2 l_p^2)^{2d}\frac{1}{A^d}]=-4\pi
G(\rho+p),\label{fr3}\\
\frac{8\pi
G}{3}\rho=4\pi[\frac{1}{A}+\frac{1}{2}\alpha^2l_p^2\frac{1}{A^2}+\frac{2}{3}(\pi\alpha^2l_p^2)^2\frac{1}{A^3}\nonumber
\\
+\sum_{d=3}\frac{c_d}{d+1}(4\pi\alpha^2l_p^2)^{2d}\frac{1}{A^{d+1}}].\label{fr4}
\end{eqnarray}

\subsection{$n>3$ and $n$ is an odd number}
When $n$ is an odd number, substituting (\ref{f}) into (\ref{area3})
and integrating, we have
\begin{eqnarray}
A_G&=&A+\sum_{d=1}^{d=\frac{n-3}{2}}c_d(\alpha l_p)^{2d}\frac{n-1}{n-2d-1}A(\frac{n\Omega_n}{A})^{\frac{2d}{n-1}}\nonumber\\
&+&c_{\frac{n-1}{2}}(\alpha l_p)^{n-1}n\Omega_nlnA\nonumber\\
&+&\sum_{d=\frac{n+1}{2}}c_d(\alpha
l_p)^{2d}\frac{n-1}{n-2d-1}A(\frac{n\Omega_n}{A})^{\frac{2d}{n-1}}.
\end{eqnarray}
By making use of Bekenstein-Hawking area law, $S=A/4G$, we can
obtain the expression of the entropy of the apparent horizon after
taking into account the effect of GUP. That is, the correction to
entropy is given by
\begin{eqnarray}
S_G&=&\frac{A}{4G}+\sum_{d=1}^{d=\frac{n-3}{2}}c_d(\alpha
l_p)^{2d}\frac{n-1}{n-2d-1}\frac{A}{4G}(\frac{n\Omega_n}{A})^{\frac{2d}{n-1}}\nonumber\\
&+&c_{\frac{n-1}{2}}\frac{(\alpha l_p)^{n-1}}{4G}n\Omega_nln\frac{A}{4G}\nonumber\\
&+&\sum_{d=\frac{n+1}{2}}c_d(\alpha
l_p)^{2d}\frac{n-1}{n-2d-1}\frac{A}{4G}(\frac{n\Omega_n}{A})^{\frac{2d}{n-1}}\nonumber\\
&+&const.\label{entropy2}
\end{eqnarray}
It is obvious that the logarithmic correction term exists when $n$
is an odd number.

In order to obtain the modified Friedmann equations from the
modified entropy-area relation (\ref{entropy2}) for
$(n+1)$-dimensional FRW spacetime, we have to generalize Cai, Cao
and Hu's approach to a $(n+1)$-dimensional FRW universe, while the
original approach in \cite{cao2} is only valid in
($3+1$)-dimensional FRW universe. The generalization is simple. The
first law of thermodynamics on the apparent horizon $dE=TdS$ leads
to
\begin{eqnarray}
A(\rho+p)H\tilde{r}_Adt=\frac{1}{2\pi \tilde{r}_A}dS_G,\label{first
law}
\end{eqnarray}
here $A(\rho+p)H\tilde{r}_Adt=dE$ is the amount of energy having
crossed the apparent horizon. By way of some simple manipulations,
we can obtain the Friedmann equations in ($n+1$)-dimensional FRW
universe, that is
\begin{eqnarray}
(\dot{H}-\frac{k}{a^2})f_G(A)=-\frac{8\pi
G}{n-1}(\rho+p),\label{fridemann3}\\
\frac{8\pi G}{n}\rho=-\int
f_G(A)(\frac{A}{n\Omega_n})^{\frac{-2}{n-1}}\frac{dA}{A},\label{fridemann4}
\end{eqnarray}
Substituting (\ref{f}) into (\ref{fridemann3}) and
(\ref{fridemann4}), we can obtain the modified Friedmann equations
in ($n+1$)-dimensional FRW spacetime including the consideration of
the GUP,
\begin{eqnarray}
(\dot{H}-\frac{k}{a^2})[1+\frac{\alpha^2l_p^2}{4}(\frac{n\Omega_n}{A})^{\frac{2}{n-1}}+
\frac{(\alpha^2l_p^2)^2}{8}(\frac{n\Omega_n}{A})^{\frac{4}{n-1}}\nonumber\\
+\sum_{d=3}c_d(\alpha
l_p)^{2d}(\frac{n\Omega_n}{A})^{\frac{2d}{n-1}}]=-\frac{8\pi
G}{n-1}(\rho+p),\label{fridemann5}\\
\frac{16\pi
G}{n(n-1)}\rho=(\frac{n\Omega_n}{A})^{\frac{2}{n-1}}\nonumber\\
+ \sum_{d=1}\frac{c_d}{d+1}(\alpha
l_p)^{2d}(\frac{n\Omega_n}{A})^{\frac{2d+2}{n-1}}.\label{fridemann6}
\end{eqnarray}
We note here that the above equations are independent on whether $n$
is an odd or even number. When we take $n=3$, Eq.(\ref{fridemann5})
and (\ref{fridemann6}) reduce to Eq.(\ref{fr3}) and (\ref{fr4})
respectively.

\subsection{$n>3$ and $n$ is an even number}
When $n$ is an even number, following the same route above, we can
obtain the expression of the entropy of the apparent horizon after
taking into account the effect of GUP, which is
\begin{eqnarray}
S_G&=&\frac{A}{4G}+\frac{\alpha^2l_p^2}{4}\frac{n-1}{n-3}\frac{A}{4G}(\frac{n\Omega_n}{A})^{\frac{2}{n-1}}\nonumber\\
&+&\sum_{d=2}c_d(\alpha
l_p)^{2d}\frac{n-1}{n-2d-1}\frac{A}{4G}(\frac{n\Omega_n}{A})^{\frac{2d}{n-1}}.\label{entr}
\end{eqnarray}
From this expression, when $d$ is an even number, the logarithmic
term does not exist in the correction to the entropy of the apparent
horizon of FRW spacetime. This implies that the logarithmic
correction term in the entropy of the apparent horizon is
dimensional dependent.

Since the derivation of the modified Friedmann equations
(\ref{fridemann5},\ref{fridemann6}) in ($n+1$)-dimensional FRW
universe is not relevant to that whether $n$ is an even or odd
number, (\ref{fridemann5},\ref{fridemann6}) are also valid when $n$
is an even. Therefore, the modified Friedmann equations from the
modified entropy (\ref{entr}) are just
Eqs.(\ref{fridemann5},\ref{fridemann6}).

\section{The EUP case}
The GUP is the high energy correction to the conventional Heisenberg
uncertainty relation. In large length scales, the GUP is
unimportant. In this case, one might consider an extension of the
uncertainty relation which contains the effect of the large length
scales. The extended uncertainty principle is given
by\cite{gup-hawking1,eup}
\begin{eqnarray}
\delta x\delta p \geq 1+\beta^2 \frac{\delta x^2}{l^2},\label{eup}
\end{eqnarray}
where $\beta$ is a dimensionless real constant, and $l$ is an
unknown fundamental characteristic large length scale. The EUP
implies that there is a minimal momentum
\begin{eqnarray}
\delta p\geq \frac{1}{\delta x}+\frac{\beta^2}{l^2}\delta x \geq
2|\frac{\beta}{l}|.
\end{eqnarray}

From EUP (\ref{eup}), the uncertainty of momentum can be written as
\begin{eqnarray}
\delta p \geq \frac{1}{\delta x}+\frac{\beta^2}{l^2}\delta
x=\frac{1}{\delta x}f_E(\delta x),\label{eup2}
\end{eqnarray}
where
\begin{eqnarray}
f_E(\delta x)=1+\frac{\beta^2\delta x^2}{l^2}
\end{eqnarray}
is the departure function in EUP case.

Now, with the same approach in Section II, it is easy to obtain the
entropy of the apparent horizon after taking into account the effect
of EUP,
\begin{eqnarray}
S_E=\frac{A}{4G}+\frac{4\beta^2}{
l^2}\frac{n-1}{n+1}(\frac{A}{n\Omega_n})^{\frac{2}{n-1}}\frac{A}{4G}.\label{entropy9}
\end{eqnarray}
The corresponding modified Friedmann equations are expressed as
\begin{eqnarray}
(\dot{H}-\frac{k}{a^2})(1+\frac{4\beta^2}{
l^2}(\frac{A}{n\Omega_n})^{\frac{2}{n-1}})=-\frac{8\pi
G}{n-1}(\rho+p),\nonumber\\
\frac{16\pi
G}{n(n-1)}\rho=(\frac{n\Omega_n}{A})^{\frac{2}{n-1}}+\frac{4\beta^2}{
l^2}ln A.
\end{eqnarray}
When $n=3$, the modified Friedmann equations are
\begin{eqnarray}
(\dot{H}-\frac{k}{a^2})(1+\frac{\beta^2}{\pi l^2}A)=-4\pi
G(\rho+p),\nonumber\\
\frac{8\pi G}{3}\rho=\frac{4\pi}{A}+\frac{4\beta^2}{l^2}lnA.
\end{eqnarray}

Here are some remarks: First, in the derivation of the modified
entropy (\ref{entropy9}), we didn't impose any limit on the
inequality (\ref{eup2}) as in (\ref{f}). This means that the entropy
(\ref{entropy9})is an exact expression. Second, although the
derivation of entropy (\ref{entropy9}) is in the context of the FRW
universe, it is very easy to generalize it to the black hole
physics. For example, with the similar procedure, one can easily
find that the entropy (\ref{entropy9}) is also valid in
Schwarzschild black hole after the consideration of the effect of
EUP.

Third, it is of interest to consider a more general case that
combine both the GUP and the EUP and is named the generalized
extended uncertainty principle (GEUP). The GEUP is given
by\cite{gup-hawking1,eup1}
\begin{eqnarray}
\delta x \delta p\geq 1+\alpha^2l_p^2\delta p^2+\beta^2\frac{\delta
x^2}{l^2}.
\end{eqnarray}
From this expression, it is easy to obtain the uncertainty of
momentum
\begin{eqnarray}
\delta p\geq \frac{1}{\delta x}f_{GE}(\delta x^2),
\end{eqnarray}
where
\begin{eqnarray}
f_{GE}(\delta x^2)=\frac{\delta x^2}{2\alpha^2
l_p^2}[1-\sqrt{1-\frac{4\alpha^2l_p^2}{\delta
x^2}[1+\beta^2\frac{\delta x^2}{l^2}]}]
\end{eqnarray}
is the departure function in the case of the GEUP. By closely
following the procedure in Section II, one can also obtain a
modified entropy of the apparent horizon of the FRW universe and the
corresponding modified Friedmann equations including the effect of
the GEUP.

\section{Summary and Discussions}
In this paper, we have investigated the influence of the GUP and the
EUP on the thermodynamics of the FRW universe. We have shown that
the GUP and EUP contribute corrections to the conventional
entropy-area relation on the apparent horizon of the FRW universe as
well as the Friedmann equations. The later implies that the GUP and
EUP can influent the dynamics of the FRW universe. In particular, in
the case of the GUP, we have shown that the leading logarithmic
correction term exists only for the (odd number+$1$) dimensional FRW
spacetime, and moreover, the leading logarithmic term gives a
positive contribution to the entropy of the apparent horizon. For
(even number+$1$) dimensional FRW spacetime, there is not a
logarithmic correction term in the entropy of the apparent horizon.

It is worthwhile to point out that the results in this paper can be
generalized in some ways. First, it is of interest to search the
statistics meaning of the modified entropy on the apparent horizon
of the FRW universe. In \cite{brick}, using the brick wall method,
the authors have calculated the statistics entropy of a scalar field
in FRW universe. How can one modify their results if one take into
account of the effect of the GUP or EUP? This is an interesting
problem and it needs further investigation.

Second, we have known that the GUP and EUP can modify the dynamics
of the universe. In the early universe, the high energy effects may
be important. This means the GUP may play an important role in the
early time of our universe. On the other hand, as a large length
scale effect, the EUP may be important in the late time universe.
Does the modification of the GUP or the EUP to the Friedmann
equations have some observational effects? How can one probe them?
These problems will be considered in our further workings.

Third, the GUP and the EUP modify the thermodynamics of both black
holes and FRW universe. More generally, it is of great interest to
investigate the influence of the GUP or EUP on the Einstein
equation. The investigation in this direction may provide a deeper
insight into the understanding of the quantum gravity or large
length scale corrections to the classical Einstein gravity.
\begin{acknowledgments}
This work was supported by the National Natural Science Foundation
of China (Grant No. 10275030) and Cuiying Project of Lanzhou
University (Grant No. 225000-582404).

 \end{acknowledgments}


\begin{thebibliography}{99} \addcontentsline{toc}{section}{Bibliography}
\bibitem{Black Hole}J. M. Bardeen, B. Carter, and S. W. Hawking, Commun.
Math. Phys. 31, 161 (1973); S. W. Hawking, Commun. Math. Phys. 43
(1975) 199; 46 (1976) 206(E); J. D. Bekenstein, Phys. Rev. D 7
(1973) 2333.

\bibitem{ql1}S. A. Hayward, Phys. Rev. D. 49, 6467 (1994); Class. Quant. Grav. 11,
3025 (1994) [arXiv: gr-qc/9406033]; Phys. Rev. Lett. 93 (2004)
251101 [arXiv: gr-qc/0404077]; Phys. Rev. D 70 (2004) 104027
[arXiv:gr-qc/0408008]; Phys. Rev. D 53 (1996) 1938
[arXiv:gr-qc/9408002]; Class. Quant. Grav. 15 (1998) 3147
[arXiv:gr-qc/9710089]; S. Mukohyama and S. A. Hayward, Class. Quant.
Grav. 17 (2000) 2153 [arXiv:gr-qc/9905085]; Ji-Rong Ren and Ran Li,
arXiv:0705.4339.

\bibitem{Jacobson}T. Jacobson, Phys. Rev. Lett. 75, 1260 (1995); A.
Pesci, Class. Quant.Grav. 24 (2007) 6219[arXiv: 0708.3729].

\bibitem{gauss-love1}R. G. Cai and S. P. Kim, JHEP 02 (2005) 050.

\bibitem{fr}M. Akbar and R. G. Cai, Phys. Lett. B 648, 243 (2007)
[arXiv:gr-qc/0612089]; M. Akbar and R. G. Cai, Phys. Lett. B 635
(2006) 7[arXiv:hep-th/0602156]; C. Eling, R. Guedens, and T.
Jacobson, Phys. Rev. Lett. 96 (2006) 121301 [arXiv:gr-qc/0602001].

\bibitem{gb}R. G. Cai, L. M. Cao, Y. P. Hu, and S. P. Kim,
arXiv:0810.2610; R. G. Cai and L. M. Cao, Phys. Rev. D 75 (2007)
064008 [arXiv:gr-qc/0611071]; M. Akbar and R. G. Cai, Phys. Rev. D
75, 084003 (2007) [arXiv:hep-th/0609128]; A. Paranjape, S. Sarkar,
T. Padmanabhan, Phys. Rev. D 74 (2006) 104015
[arXiv:hep-th/0607240].

\bibitem{brane}R. G. Cai and L. M. Cao, Nucl. Phys. B 785 (2007)
135[arXiv:hep-th/0612144]; A. Sheykhi, B. Wang, and R. G. Cai, Nucl.
Phys. B 779 (2007) 1 [arXiv:hep-th/0701198]; X. H. Ge, Phys. Lett. B
651 (2007) 49 [arXiv:hep-th/0703253]; R. G. Cai, Prog. Theore. Phys.
Suppl. No. 172, 100 (2008) [arXiv:0712.2142]; A. Sheykhi, B. Wang,
and R. G. Cai, Phys. Rev. D 76, 023515 (2007)
[arXiv:hep-th/0701261]; T. Zhu, J. R. Ren, and S. F. Mo,
arXiv:0805.1162.

\bibitem{padm}T. Padmanabhan, Class. Quant. Grav. 19 (2002)
5387[arXiv: gr-qc/0204019]; T. Padmanabhan, Astrophys.Space Sci. 285
(2003) 407 [arXiv:gr-qc/0209088]; T. Padmanabhan, Phys. Rept. 406
(2005) 49[arXiv:gr-qc/0311036]; T. Padmanabhan, Int. J. Mod. Phys.
D15 (2006) 1659-1676 [arXiv:gr-qc/0606061]; S. Sarkar and T.
Padmanabhan, Entropy 9 (2007) 100-107 [arXiv:gr-qc/0607042]; D.
Kothawala, S. Sarkar, and T. Padmanabhan, arXiv:gr-qc/0701002; T.
Padmanabhan and A. Paranjape, Phys.Rev. D75 (2007) 064004
[arXiv:gr-qc/0701003].

\bibitem{ql3}M. Nozawa and H. Maeda, Class. Quant. Grav. 25, 055009
(2008) [arXiv:0710.2709]; H. Maeda and M. Nozawa, Phys. Rev. D 77,
064031 (2008) [arXiv:0709.1199].

\bibitem{mass}Y. Gong and A. Wang, Phys. Rev. Lett. 99 (2007) 211301
[arXiv:0704.0793]; S. F. Wu, G. H. Yang, and P. M. Zhang,
arXiv:0710.5394; S. F. Wu, B. Wang, and G. H. Yang, Nucl. Phys. B
799 (2008) 330[arXiv:0711.1209]; S. F. Wu, G. H. Yang, and P. M.
Zhang, Prog. Theor. Phys. 120 (2008) 615[arXiv:0805.4044]; M. Akbar,
arXiv:0808.0169; E. Elizalde and P. J. Silva, arxiv: 0804.3721.

\bibitem{cft/frw}E. Papantonopoulos and V. Zamarias, JCAP 0404 (2004)
004 [arXiv: hep-th/0307144]; E. Verlinde, arXiv:hep-th/0008140.

\bibitem{brick}W. Kim, E. J. Son, and M. Yoon, arXiv:0808.1805.

\bibitem{cao2}R. G. Cai, L. M. Cao, and Y. P. Hu, JHEP 08 (2008) 090
[arXiv:0807.1232].

\bibitem{hawking}S. W. Hawking, Commun. Math. Phys. 43 (1975) 199; S. W.
Hawking, Phys. Rev. D 14 (1976) 2460; M. K. Parikh and F. Wilczek,
Phys. Rev. Lett. 85 (2000) 5042-5045 [arXiv:hep-th/9907001].

\bibitem{cao1}R. G. Cai, L. M. Cao, and Y. P. Hu, arXiv:0809.1554.

\bibitem{gup1}A. Kempf, G. Mangano, and R. B. Mann, Phys. Rev. D 52 (1995) 1108-1118
[arXiv:hep-th/9412167].

\bibitem{gup2}M. Maggiore, Phys. Lett. B 319 (1993) 83 [arXiv:
hep-th/9309034]; M. Maggiore, Phys. Rev. D 49 (1994) 5182-5187
[arXiv:hep-th/9305163].

\bibitem{gup3-string}D. Amati, M. Ciafaloni, and G. Veneziano, Phys.
Lett. B 216 (1989) 41; Nucl. Phys. B 347 (1990) 550; Nucl. Phys. B
403 (1993) 707; K. Konishi, G. Paffuti, and P. Provero, Phys. Lett.
B 234 (1990) 276.

\bibitem{gup-hawking}R. J. Adler, P. Chen, and D. I. Santiago,
Gen.Rel.Grav. 33 (2001) 2101-2108 [arXiv:gr-qc/0106080]; K. Nozari
and S. H. Mehdipour, Mod. Phys. Lett. A 20 (2005) 2937
[arXiv:0809.3144].

\bibitem{gup-hawking1}M. Park,
Phys. Lett. B 659 (2008) 698-702 [arXiv:0709.2307].

\bibitem{gup-entropy1}K. Nozari and A. S. Sefiedgar, Gen. Rel. Grav. 39 (2007) 501-509
[ arXiv:gr-qc/0606046]; A. J. M. Medved and E. C. Vagenas, Phys.
Rev. D 70 (2004) 124021 [arXiv:hep-th/0411022]; R. Zhao and S. L.
Zhang, Phys. Lett. B 641 (2006) 208-211; H. X. Zhao, H. F. Li, S. Q.
Hu, and R. Zhao, arXiv:gr-qc/0608023.

\bibitem{gup-entropy2}R. Zhao, Y. Q. Wu, and L. C. Zhang, Class. Quant.
Grav. 20 (2003) 4885; X. Li, Phys. Lett. B 540 (2002) 9-13
[arXiv:gr-qc/0204029]; W. Kim, Y. W. Kim, and Y. J. Park, Phys. Rev.
D 74 (2006) 104001 [arXiv:gr-qc/0605084]; Y. S. Myung, Y. W. Kim,
and Y. J. Park, Phys. Lett. B 645 (2007) 393-397
[arXiv:gr-qc/0609031]; W. Kim, and J. J. Oh, JHEP 0801 (2008) 034
[arXiv:0709.0581]; W. Kim, E. J. Son, and M. Yoon, JHEP 01 (2008)
035 [arXiv:0711.0786]; Zhao Ren, Zhang Li-Chun, Li Huai-Fan, and Wu
Yue-Qin, arXiv:0801.4118.

\bibitem{eup}C. Bambi and F. R. Urban, Class. Quant. Grav. 25 (2008)
095006 [arXiv:0709.1965].

\bibitem{eup1}X. Han, H. Li, and Y. Ling, Phys. Lett. B 666 (2008)
121 [arXiv:0807.4269]; B. Bolen and M. Cavaglia, Gen.Rel.Grav. 37
(2005) 1255-1262 [arXiv:gr-qc/0411086]; I. Arraut, D. Batic, and M.
Nowakowski, arXiv:0810.5156.

\bibitem{log1}R. K. Kaul and P. Majumdar, Phys. Rev. Lett. 84, 5255
(2000) [arXiv: gr-gc/0002040]; A. Ghosh and P. Mitra, Phys. Rev. D
71, 027502 (2005) [arXiv: gr-qc/0401070]; M. Domagala and J.
Lewandowski, Class. Quant. Grav. 21, 5233 (2004) [arXiv:
gr-qc/0407051]; K. A. Meissner, Class. Quant. Grav. 21, 5245 (2004)
[arXiv: gr-qc/0407052].

\bibitem{log2}A. J. M. Medved, Class.Quant.Grav. 22 (2005) 133-142
[arXiv:gr-qc/0406044].

\bibitem{other}M. R. Setare, Phys. Rev. D 70 (2004) 087501
[arXiv:hep-th/0410044]; Int. J. Mod. Phys. A 21 (2006) 1325
[arXiv:hep-th/0504179]; S. Das and E. C. Vagenas, Phys. Rev. Lett.
101 (2008) 221301[arXiv:0810.5333]; M. V. Battisti and G. Montani,
Phys. Lett. B 656 (2007) 96 [arXiv:gr-qc/0703025]; Phys. Rev. D 77
(2008) 023518[arXiv:0707.2726]; arXiv:0808.0831; M. V. Battisti,
arXiv:0805.1178; S. F. Hassan and M. S. Sloth, Nucl. Phys. B 674
(2003) 434[arXiv:hep-th/0204110].

\bibitem{other1}R. Banerjee and B. R. Majhi, Phys. Lett. B 662 (2008)
62 [arXiv:0801.0200]; JHEP 0806 (2008) 095 [arXiv:0805.2220];
arXiv:0808.3688.
\end{thebibliography}
\end{document}